\documentclass[twocolumn,conference, 10pt]{IEEEtran}
\usepackage[latin9]{inputenc}
\usepackage{float}
\usepackage{bm}
\usepackage{amsmath}
\usepackage{amsthm}
\usepackage{amssymb}
\usepackage{graphicx}

\makeatletter

\floatstyle{ruled}
\newfloat{algorithm}{tbp}{loa}
\providecommand{\algorithmname}{Algorithm}
\floatname{algorithm}{\protect\algorithmname}

\theoremstyle{plain}
\newtheorem{thm}{\protect\theoremname}
\theoremstyle{plain}
\newtheorem{prop}[thm]{\protect\propositionname}

\usepackage[caption=false,font=footnotesize]{subfig}
\usepackage{algorithmic}
\usepackage{algorithm}
\allowdisplaybreaks
\usepackage{flushend}

\makeatother

\providecommand{\propositionname}{Proposition}
\providecommand{\theoremname}{Theorem}

\begin{document}

\title{Wideband Massive MIMO Channel Estimation via Sequential Atomic Norm
Minimization}

\author{\IEEEauthorblockN{Stelios~Stefanatos\IEEEauthorrefmark{1}, Mahdi Barzegar Khalilsarai\IEEEauthorrefmark{2},
and Gerhard~Wunder\IEEEauthorrefmark{1}}\IEEEauthorblockA{\IEEEauthorrefmark{1}Heisenberg Communications and Information Theory
Group, Freie Universität Berlin, Germany}\IEEEauthorblockA{\IEEEauthorrefmark{2}Communications and Information Theory Group,
Technische Universität Berlin, Germany\\
Email: stelios.stefanatos@fu-berlin.de, m.barzegarkhalilsarai@tu-berlin.de,
g.wunder@fu-berlin.de}}
\maketitle
\begin{abstract}
The recently introduced atomic norm minimization (ANM) framework for
parameter estimation is a promising candidate towards low-overhead
channel estimation in wireless communications. However, previous works
on ANM-based channel estimation evaluated performance on channels
with artificially imposed channel path separability, which cannot
be guaranteed in practice. In addition, direct application of the
ANM framework for massive MIMO channel estimation is computationally
infeasible due to the large signal dimensions. In this paper, a low-complexity
ANM-based channel estimator for wideband massive MIMO is proposed,
consisting of two sequential steps, the first step estimating the
channel over the angle-of-arival/spatial domain and the second step
over the delay/frequency domain. The mean squared error performance
of the proposed estimator is analytically characterized in terms of
tight lower bounds. It is shown that the proposed algorithm achieves
excellent performance that is close to the best that can be achieved
by any unbiased channel estimator in the regime of low to moderate
number of channel paths, without any restrictions on their separability.
\end{abstract}

\IEEEpeerreviewmaketitle{}

\section{Introduction}

The extremely large number of signal dimensions available in massive
MIMO offer significant performance benefits, however, they also come
at a significant cost related to the overhead required for channel
estimation \cite{Marzetta mMIMO}. Towards reducing this cost, a recent
line of works explores the fundamentally sparse parametric representation
of the wireless propagation channel \cite{Sayeed MIMO channel model}
and employs concepts and algorithms from the field of compressive
sensing (CS) towards estimating the channel parameters with minimal
overhead \cite{CS for channel est Mag}.

One of the most recent tools in CS is that of atomic norm minimization
(ANM), which deals with the estimation of superimposed harmonic signals
from a minimal number of (noisy) samples \cite{Candes,CS off the grid}.
Rigorous performance guarantees for this task are available, based
on the condition that the constituent signal frequencies are sufficiently
separated \cite{CS off the grid}. Realizing that the massive MIMO
scenario with single antenna transmitters and narrow band signaling
fits this model, \cite{ICC 2015 ANM mMIMO} proposed an ANM-based
channel estimator. Significant performance gains over conventional
channel estimation approaches were demonstrated, however, in a scenario
where the channel paths were (artificially) imposed to be sufficiently
separated in the angle-of-arrival (spatial) domain. When wideband
(OFDM) massive MIMO transmissions are considered, the channel can
be treated as a superposition of two-dimensional harmonic signals,
for which the ANM framework has recently been applied \cite{ANM 2D Chi,Stoica Vandermonde decomp},
again with performance guarantees under sufficient separation of the
(two-dimensional) frequencies. Unfortunately, due to the large receive
antenna arrays and number of subcarriers, direct application of the
two-dimensional ANM algorithm is computationally infeasible. In \cite{ICC 2017 truncated ANM for mMIMO},
a, so-called, truncated ANM approach was proposed. However, this approach
works only for a specific set of pilot subcarriers, which limits its
applicability to one user estimated per OFDM symbol, and with results
demonstrated only for channels with $2$ paths (thus guaranteeing
with high probability sufficiently separated paths). A heuristic,
low-complexity algorithm for two-dimensional ANM was independently
proposed in \cite{ICASP 2017 second,ICASP 2017}. However, performance
of the algorithm was only evaluated with sufficiently separable channel
paths, as the algorithm fails otherwise.

In this paper, the problem of uplink wideband massive MIMO channel
estimation is considered via an ANM-based approach that is both low-complexity
\emph{and }robust to channel paths separation. Towards this end, a
two-step algorithm is proposed consisting of two sequential multiple
measurement vectors (MMV) ANM problems, estimating the channel matrix,
first over the angle-of-arrival/spatial and then over the delay/frequency
domain. Decoupling the estimation of the domains significantly reduces
complexity as well as relaxes the performance sensitivity on path
separability on the angle-delay plane. A tight analytical bound for
the mean-squared-error performance of the proposed algorithm is derived,
which, in addition to numerical evidence, indicates that, in a setting
with a small-to-moderate number of channel paths (and no artificial
path separability imposed), the proposed algorithm significantly outperforms
previous approaches and has a performance close to the best that can
be achieved by any unbiased estimator.

\emph{Notation}: Vectors (matrices) will be denoted by lower (upper)
case bold letters. The set $\{0,1,\ldots,N-1\}$ will be denoted as
$[N]$. $\mathbf{X}^{T}$, $\mathbf{X}^{H}$, $\text{tr}\{\mathbf{X}\}$,
and $\lVert\mathbf{X}\rVert\triangleq\sqrt{\text{tr}\{\mathbf{X}^{H}\mathbf{X}\}}$
denote the transpose, Hermitian, trace, and Frobenius norm of $\mathbf{X}$,
respectively. $\mathbf{X}\succeq\mathbf{0}$ means that $\mathbf{X}$
is positive semidefinite. The cardinality of a set $\mathcal{A}$
is denoted by $|\mathcal{A}|$. The complex exponential sequence of
length $K$ is denoted as $\mathbf{f}_{K}(\phi)\triangleq[1,e^{-i2\pi\phi},\ldots,e^{-i2\pi\phi(K-1)}]^{T}$,
$\phi\in[0,1]$.

\section{System Model and Channel Estimation via Atomic Norm Minimization}

\subsection{System Model}

The uplink of a single cell, wideband, massive MIMO system is considered.
The user equipments (UEs) have single antennas whereas the base station
(BS) is equipped with a uniform linear array (ULA) of $M\gg1$ antennas,
enumerated by the set $[M]$. Transmissions are performed via OFDM
with $N\gg1$ subcarriers, enumerated by the set $[N]$. Considering
an arbitrary UE, its complex baseband channel transfer matrix $\mathbf{H}\in\mathbb{C}^{M\times N}$,
whose $(m,n)$-th element corresponds to the channel gain at antenna
$m\in[M]$ and subcarrier $n\in[N]$, equals (after appropriate normalizations)
\cite{ICASP 2017,pilot decontamination wideband mMIMO}
\begin{equation}
\mathbf{H}=\sum_{l=0}^{L-1}c_{l}\mathbf{f}_{M}(\theta_{l})\mathbf{f}_{N}^{H}(\tau_{l}),\label{eq:channel model}
\end{equation}
where $L\ll M,N$ is the number of channel (propagation) paths and
$c_{l}\in\mathbb{C}$, $\theta_{l}\in[0,1]$, $\tau_{l}\in[0,1]$
are the gain, angle of arrival (AoA), and delay of the $l$-th channel
path, respectively. No \emph{a priori} assumptions on the statistics
of the channel path parameters are considered, except for the technical
assumption that two paths cannot have the same delay and AoA (otherwise,
their superposition would be equivalent to a single path). The value
of $L$ is assumed known at the BS.

Towards estimating $\mathbf{H}$ at the BS, a set $\mathcal{N}_{p}\subseteq[N]$
of $N_{p}\triangleq|\mathcal{N}_{p}|$ subcarriers is employed for
transmission of pilot symbols. In addition, only the observations
from a set $\mathcal{M}_{p}\subseteq[M]$ of $M_{p}\triangleq|\mathcal{M}_{p}|$
antenna elements is considered at the BS for implementation complexity
reduction \cite{Haghighatshoar Low dim proj}. Although $\mathcal{N}_{p}$
and $\mathcal{M}_{p}$ can, in principle, be optimized according to
some appropriate criterion, they will be both assumed in this work
as \emph{randomly and uniformly selected} from $[N]$ and $[M]$,
respectively. This approach results in a robust design (i.e., not
tied to a specific criterion and UE) and also leads to a tractable
analysis. However, $N_{p}$ and $M_{p}$ are treated as design parameters,
which, ideally, should be as small as possible.

Assuming, without loss of generality, that all pilot symbols are equal
to $1$, the BS task is to infer $\mathbf{H}$ from the $M_{p}\times N_{p}$
space-frequency observation matrix \cite{pilot decontamination wideband mMIMO}
\begin{equation}
\mathbf{Y}=\mathbf{S}_{\mathcal{M}_{p}}\mathbf{H}\mathbf{S}_{\mathcal{N}_{p}}^{T}+\mathbf{Z},\label{eq: observations}
\end{equation}

\noindent where $\mathbf{S}_{\mathcal{M}_{p}}\in\{0,1\}^{M_{p}\times M}$
, $\mathbf{S}_{\mathcal{N}_{p}}\in\{0,1\}^{N_{p}\times N}$ are downsampling
matrices, extracting the rows corresponding to the sets $\mathcal{M}_{p}$,
$\mathcal{N}_{p}$, respectively, and $\mathbf{Z}\in\mathbb{C}^{M_{p}\times N_{p}}$
is a noise matrix of independent, identically distributed (i.i.d.)
complex Gaussian elements of zero mean and variance $\sigma^{2}$. 

As $\mathbf{H}$ is a function of the path parameters $\{(c_{l},\theta_{l},\tau_{l})\}_{l\in[L]}$,
its maximum likelihood (ML) estimate can be obtained from the ML estimate
of the path parameters \cite{Kay Estimation theory}. However, the
non-linear dependence of $\mathbf{H}$ on $\{(\theta_{l},\tau_{l})\}_{l\in[L]}$
makes the (exact) computation of the ML estimates of the latter extremely
difficult for $L\geq2$ \cite{Stoica book} and motivates consideration
of suboptimal but lower-complexity alternatives for the estimation
of $\mathbf{H}$. 

\subsection{Channel Estimation via Atomic Norm Minimization}

Towards an efficient channel estimation algorithm, it is essential
to take into account the well-defined structure of $\mathbf{H}$ as
specified by (\ref{eq:channel model}). In particular, note that $\mathbf{H}$
is a (low) rank-$L$ matrix and, also, $\mathbf{H}\in\text{span}(\mathcal{A})$,
where 
\[
\mathcal{A}\triangleq\left\{ \mathbf{f}_{M}(\theta)\mathbf{f}_{N}^{H}(\tau):(\theta,\tau)\in[0,1]^{2}\right\} ,
\]

\noindent is an uncountable \emph{atom set}. Defining the \emph{atomic
norm} of an arbitrary matrix $\mathbf{X}\in\mathbb{C}^{M\times N}$
with respect to (w.r.t.) a set $\mathcal{U}$ with elements (atoms)
$\mathbf{U}(\chi)$, where $\chi$ is a continuous-valued index, as
\cite{linear inverse problems}
\begin{equation}
\left\Vert \mathbf{X}\right\Vert _{\mathcal{U}}\triangleq\underset{s_{k},\chi_{k}}{\text{inf}}\left\{ \sum_{k}|s_{k}|\left|\mathbf{X}=\sum_{k}s_{k}\mathbf{U}(\chi_{k}),\mathbf{U}(\chi_{k})\in\mathcal{U}\right.\right\} ,\label{eq:AN definition}
\end{equation}
an estimate of $\mathbf{H}$ can be obtained via the atomic norm minimization
(ANM) principle \cite{Candes,CS off the grid}, i.e., 
\begin{equation}
\hat{\mathbf{H}}=\underset{\mathbf{X}\in\mathbb{C}^{M\times N}}{\text{argmin}}\left\{ \left\Vert \mathbf{X}\right\Vert _{\mathcal{A}}\left|\|\mathbf{Y}-\mathbf{S}_{\mathcal{M}_{p}}\mathbf{X}\mathbf{S}_{\mathcal{N}_{p}}^{T}\|\leq\|\mathbf{Z}\|\right.\right\} .\label{eq:2D ANM problem formulation}
\end{equation}
In practice, an estimate of $\|\mathbf{Z}\|$ is used in (\ref{eq:2D ANM problem formulation})
as its exact value is not available \cite{MathIntroToCS}. Note that
the ANM approach directly provides an estimate of $\mathbf{H},$ without
explicit computation of the path parameters. However, improved performance
is achieved by (a) explicitly estimating $\{(\theta_{l},\tau_{l})\}_{l\in[L]}$
from the $L$ strongest two-dimensional harmonic components of $\hat{\mathbf{H}}$
(e.g., using standard harmonic retrieval techniques as in \cite{matrix pencil}),
(b) estimating the path gains $\{c_{l}\}_{l\in[L]}$ by solving a
simple linear least squares problem, and (c) constructing a new, \emph{denoised}
channel estimate from the estimated path parameters \cite{Recht denoising}.

A strong motivation for the ANM-based channel estimation is that,
in the noiseless case, perfect channel recovery (i.e., $\hat{\mathbf{H}}=\mathbf{H}$)
can be guaranteed with overwhelmingly high probability as long as
$M_{p}$, $N_{p}$ are sufficiently large and it holds \cite{ANM 2D Chi,Stoica Vandermonde decomp}
\begin{equation}
\underset{l\neq l'}{\text{min }}\text{max}\left\{ |\theta_{l}-\theta_{l'}|,|\tau_{l}-\tau_{l'}|\right\} >d,\label{eq:2D separability condition}
\end{equation}
where $|\cdot|$ is the wraparound distance metric on the unit circle,
and $d>0$ is a constant that restricts how close two channel paths
can lie on the delay-AoA plane. Although the necessary and sufficient
value of $d$ is not known, numerical evidence suggests that it is
approximately equal to $1/\min\left\{ M,N\right\} $. This recovery
guarantee suggests incorporation of the ANM-based channel estimator
irrespective of whether (\ref{eq:2D separability condition}) holds
or not, since, in any case, there is no control on the channel paths
that can guarantee (\ref{eq:2D separability condition}). Unfortunately,
even though the ANM problem of (\ref{eq:2D ANM problem formulation})
affords an equivalent semidefinite programming (SDP) formulation \cite{ANM 2D Chi,Stoica Vandermonde decomp},
the corresponding solution has a complexity proportional to $MN$,
which limits the practical feasibility of this approach to values
of $M,N$ up to about $10$ \cite{ICC 2017 truncated ANM for mMIMO,ICASP 2017 second,ICASP 2017},
which are too small for wideband massive MIMO applications.

Towards a low-complexity alternative, a heuristic modification of
the SDP formulation of the ANM problem was independently proposed
in \cite{ICASP 2017 second,ICASP 2017} with a computational complexity
proportional to only $M+N$. The algorithm is also able to achieve
perfect recovery in the noiseless case, however, under more restrictive
path separability conditions and/or number of observations $N_{p},M_{p}$.
Even though this approach strikes a good balance between recovery
guarantees and complexity, its main problem is that it provides \emph{decoupled}
estimates of the AoA and delay values. In the presence of noise, a
pairing of AoA and delay values must be performed in order to obtain
a denoised channel estimate, however, it is not clear how to do so
optimally (a heuristic pairing procedure that only works for the noiseless
with perfect channel recovery case is suggested in \cite{ICASP 2017}).

\section{Channel Estimation via Sequential Atomic Norm Minimization}

Motivated by the decoupling of estimated AoAs and delays considered
in \cite{ICASP 2017 second,ICASP 2017} for complexity reduction,
a new channel estimator is proposed in this section that also results
in decoupled parameter estimates, however, it can achieve denoising
without the need of a pairing step. The proposed algorithm is based
on solving sequentially two ANM problems as follows: In the first
sep, the $M\times N_{p}$ matrix $\mathbf{H}_{1}\triangleq\mathbf{H}\mathbf{S}_{\mathcal{N}_{p}}^{T}$
is estimated from $\mathbf{Y}$, by treating the $N_{p}$ columns
of $\mathbf{H}_{1}$ as multiple measurement vectors (MMV) of an one-dimensional
harmonic sequence over the AoA domain and, in the second step, $\mathbf{H}$
is recovered from $\mathbf{\hat{H}}_{1}$, this time treating the
$M$ rows of $\mathbf{H}$ as MMV of another one-dimensional harmonic
sequence over the delay domain. Informally, the proposed algorithm
decouples the AoA/spatial and delay/frequency domains in two estimation
problems, where only the parameters of one domain are explicitly estimated
and the parameters of the other domain are subsumed into the equivalent
MMV coefficients.

To be specific, note that $\mathbf{H}_{1}$ can be written as
\begin{equation}
\mathbf{H}_{1}=\sum_{l=0}^{L-1}c_{l}\mathbf{f}_{M}(\theta_{l})\mathbf{b}_{1,l}^{H},\label{eq:H1 representation}
\end{equation}
where $\mathbf{b}_{1,l}\triangleq\mathbf{S}_{\mathcal{N}_{p}}\mathbf{f}_{N}(\tau_{l}),l\in[L]$.
By ignoring the structure of $\mathbf{b}_{1,l}$ and treating it as
an arbitrary vector in $\mathbb{C}^{N_{p}}$, the representation of
(\ref{eq:H1 representation}) corresponds exactly to an MMV observation
of a harmonic sequence with harmonics corresponding to the AoA values
$\{\theta_{l}\}_{l\in[L]}$. This allows for the incorporation of
recently proposed ANM-based approaches for the estimation of MMV matrices
with the structure of (\ref{eq:H1 representation}) \cite{MMV ANM Chi conference,MMV Exact}.
In particular, considering the atom set
\[
\mathcal{A}_{\text{MMV}_{1}}\triangleq\left\{ \mathbf{f}_{M}(\theta)\mathbf{b}_{1}^{H},\theta\in[0,1],\mathbf{b}_{1}\in\mathbb{C}^{M_{p}},\|\mathbf{b}_{1}\|^{2}=1\right\} ,
\]
and temporarily assuming noiseless observations, $\mathbf{H}_{1}$
can be estimated from $\mathbf{Y}$ as 
\begin{align}
\hat{\mathbf{H}}_{1}= & \underset{\mathbf{X}\in\mathbb{C}^{M\times N_{p}}}{\text{argmin}}\left\{ \left\Vert \mathbf{X}\right\Vert _{\mathcal{\mathcal{A}_{\text{MMV}}}_{1}}\left|\mathbf{S}_{\mathcal{M}_{p}}\mathbf{X}=\mathbf{Y}\right.\right\} ,\label{eq:ANM MMV 1}
\end{align}

\noindent with $\left\Vert \mathbf{X}\right\Vert _{\mathcal{\mathcal{A}_{\text{MMV}}}_{1}}$
efficiently computed as $\frac{1}{2}\left(\text{tr}\left\{ \mathbf{T}(\mathbf{u}_{1}^{\star})\right\} +\text{tr}\left\{ \mathbf{W}_{1}^{\star}\right\} \right)$,
where $\mathbf{T}(\mathbf{u}_{1}^{\star})$ denotes the Hermitian
Toeplitz matrix with first row $\mathbf{u}_{1}^{\star}$ and $\mathbf{u}_{1}^{\star}$,
$\mathbf{W}_{1}^{\star}$ are the (unique) solutions of the SDP \cite{MMV ANM Chi conference,MMV Exact}

\begin{equation}
\left\{ \begin{array}{cc}
\underset{\mathbf{u}_{1}\in\mathbb{C}^{M},\mathbf{W}_{1}\in\mathbb{C}^{N_{p}\times N_{p}}}{\text{minimize }} & \frac{1}{2}\left(\text{tr}\left\{ \mathbf{T}(\mathbf{u}_{1})\right\} +\text{tr}\left\{ \mathbf{W}_{1}\right\} \right)\\
\text{subject to } & \left[\begin{array}{cc}
\mathbf{T}(\mathbf{u}_{1}) & \mathbf{X}\\
\mathbf{X}^{H} & \mathbf{W}_{1}
\end{array}\right]\succeq\mathbf{0}
\end{array}\right\} .\label{eq:MMV ANM as SDP}
\end{equation}

\noindent Note that the complexity of solving the SDP of (\ref{eq:MMV ANM as SDP})
is proportional to $M+N_{p}$ (size of the matrix constraint). Perfect
recovery of $\mathbf{H}_{1}$ can also be guaranteed for this problem
with sufficiently large $N_{p}$ and separation of the AoAs $\{\theta_{l}\}_{l\in[L]}$
\cite{MMV Exact}, whereas the AoAs themselves can be recovered as
the $L$ harmonics of the Vandermonde decomposition (VD) of $\mathbf{T}(\mathbf{u}_{1}^{\star})$
\cite{Stoica DoA arxiv}.

Assuming perfect recovery of $\mathbf{H}_{1}$ and writing $\mathbf{H}$
as
\[
\mathbf{H}=\sum_{l=0}^{L-1}c_{l}\mathbf{b}_{2,l}\mathbf{f}_{N}(\tau_{l})^{H},
\]
with $\mathbf{b}_{2,l}\triangleq\mathbf{f}_{M}(\theta_{l}),l\in[L]$,
treated as an arbitrary vector in $\mathbb{C}^{M}$, $\mathbf{H}$
can be estimated from $\hat{\mathbf{H}}_{1}=\mathbf{H}\mathbf{S}_{\mathcal{N}_{p}}^{T}$
by solving a second MMV ANM problem 
\begin{equation}
\hat{\mathbf{H}}=\underset{\mathbf{X}\in\mathbb{C}^{M\times N}}{\text{argmin}}\left\{ \left\Vert \mathbf{X}\right\Vert _{\mathcal{\mathcal{A}_{\text{MMV}}}_{2}}\left|\mathbf{X}\mathbf{S}_{\mathcal{N}_{p}}^{T}=\hat{\mathbf{H}}_{1}\right.\right\} ,\label{eq:ANM MMV 2}
\end{equation}

\noindent where 
\[
\mathcal{A}_{\text{MMV}_{2}}\!\triangleq\!\left\{ \mathbf{b}_{2}\mathbf{f}_{N}^{H}(\tau),\tau\in[0,1],\mathbf{b}_{2}\in\mathbb{C}^{N},\|\mathbf{b}_{2}\|^{2}=1\right\} .
\]
Similarly to the first MMV ANM problem, computation of $\left\Vert \mathbf{X}\right\Vert _{\mathcal{\mathcal{A}_{\text{MMV}}}_{2}}$
affords an equivalent SDP formulation with a complexity proportional
to $M+N$, and with the delay values $\{\tau_{l}\}_{l\in[L]}$ extracted
from the VD of $\mathbf{T}(\mathbf{u}_{2}^{\star})$, where $\mathbf{u}_{2}^{\star}\in\mathbb{C}^{N}$
is the optimal solution variable of the SDP formulation of (\ref{eq:ANM MMV 2}).

When the observations are noisy, it is proposed to operate the algorithm
as described, i.e., by treating $\mathbf{Y}$ as noiseless, and obtaining
a denoised MMV estimate for each step of the algorithm by extracting
the AoA and delay values as the $L$ strongest harmonics of the VD
of $\mathbf{T}(\mathbf{u}_{1}^{\star})$ and $\mathbf{T}(\mathbf{u}_{2}^{\star})$,
respectively. The proposed algorithm is summarized below.

\begin{algorithm}[tbh]
\caption{\label{alg:proposed alg}Sequential MMV ANM Channel Estimation}
\begin{algorithmic}[1]
\REQUIRE $\mathbf{Y}$, $L$.
\STATE Obtain $\mathbf{u}_1^\star$ from the solution of (\ref{eq:ANM MMV 1})
\STATE Compute the VD $\mathbf{T}(\mathbf{u}_1^\star)\!=\!\sum_{k=0}^{M-1} d_{1,k} \mathbf{f}_M({\phi_{1,k}}) \mathbf{f}_M^H({\phi_{1,k}})$
\STATE $\{\hat{\theta}_l \}_{l \in [L]}$ = $\{\phi_{1,l} \}_{l \in \mathcal{L}_1}$, where $\mathcal{L}_1$ is the set of indices of the $L$ largest elements of $\{ |d_{1,k}| \}_{k\in [M]}$.
\STATE $\{ \hat{\mathbf{x}}_{1,l}\}_{l \in [L]}\! =\!\arg  \underset{\mathbf{x}_l \in \mathbb{C}^{N_p}, l \in [L]}{\min} \| \mathbf{Y} \! - \! \mathbf{S}_{\mathcal{M}_p} \sum_{l=0}^{L-1} \mathbf{f}_M(\hat{\theta}_l) \mathbf{x}_l^H      \|$
\STATE $\hat{\mathbf{H}}_1 =\sum_{l=0}^{L-1} \mathbf{f}_M(\hat{\theta}_l) \hat{\mathbf{x}}_{1,l}^H$

\STATE Solve (\ref{eq:ANM MMV 2}) and set
$\hat{\mathbf{H}} =\sum_{l=0}^{L-1} \hat{\mathbf{x}}_{2,l} \mathbf{f}_N(\hat{\tau}_l)^H$, with $\{\hat{\tau}_l, \hat{\mathbf{x}}_{2,l}\}_{l \in [L]}$ obtained following the same procedure as for $\{\hat{\theta}_l, \hat{\mathbf{x}}_{1,l}\}_{l \in [L]}$ starting from the VD of $\mathbf{T}(\mathbf{u}_2^\star)$.

\RETURN $\hat{\mathbf{H}}$
\end{algorithmic} 
\end{algorithm}
 Note that the total complexity of the proposed two-step procedure
is proportional to $2M+N+N_{p}$, which is at most two times greater
than the one of \cite{ICASP 2017 second,ICASP 2017}. In addition,
\emph{no pairing step for the estimated AoA and delay values is required
}as the algorithm essentially considers the angle and delay domains
independently. This is in contrast to \cite{ICASP 2017 second,ICASP 2017}
where the decoupled path parameters are obtained by jointly processing
the angle and delay domains.

\section{Performance Bounds}

Analytical characterization of the MSE performance of ANM-based estimators
is in general very difficult, with the only available results being
very loose upper bounds for the case of $N_{p}=N$, $M_{p}=M$ (full
observations) \cite{Chi denoising}. Towards understanding the merits
of the proposed ANM-based channel estimator, lower bounds for its
MSE performance are pursued next, starting from a universal bound
that holds for \emph{any} unbiased estimator of $\mathbf{H}$ and
\emph{without any assumptions/restrictions on the separability of
channel paths on the AoA-delay plane}, which are routinely imposed
(artificially) in related previous works \cite{ICC 2015 ANM mMIMO,ICASP 2017 second,ICASP 2017,Recht denoising,MMV ANM Chi conference,Chi denoising}.
\begin{thm}
\label{thm: 2D CRLB}The per-element MSE of any unbiased estimator
of $\mathbf{H}$ is lower bounded as
\begin{equation}
\frac{1}{MN}\mathbb{E}(\|\hat{\mathbf{H}}-\mathbf{H}\|^{2})\geq\frac{2L\sigma^{2}}{M_{p}N_{p}},\label{eq:2D CRLB bound}
\end{equation}
where the expectation is over the statistics of $\mathbf{Z}$, $\mathcal{N}_{p}$,
$\mathcal{M}_{p}$.
\end{thm}
\begin{IEEEproof}
Please see the Appendix.
\end{IEEEproof}
It is noted that for $M_{p}=M$, $N_{p}=N$, the bound of (\ref{thm: 2D CRLB})
coincides with the Cramer-Rao lower bound (CRLB) \cite{Luc}, but
it is looser otherwise (numerical experiments showed that it is very
close to the CRLB whenever $M_{p}N_{p}\geq L$). Theorem \ref{thm: 2D CRLB}
conforms to the intuition that the (minimum) MSE is proportional to
$L$ (as the number of parameters describing $\mathbf{H}$ is $3L$)
and inversely proportional to $M_{p}$ and $N_{p}$. Interestingly,
the dependence on $M_{p}$ and $N_{p}$ is only through their product
$M_{p}N_{p}$, suggesting that, for a given MSE performance, there
is flexibility in ``distributing the overhead'' to the frequency
and spatial dimensions. In particular, for a given MSE performance,
minimum pilot overhead $N_{p}$ can be achieved by utilizing all the
antenna elements, i.e., $M_{p}=M$. 

As the bound of (\ref{eq:2D CRLB bound}) is universal, i.e., it holds
for all (unbiased) estimators, a tighter MSE bound for the proposed
two-step channel estimator is pursued next, by explicitly taking into
account the treatment of the channel matrix as MMV matrices over the
AoA and delay domains.
\begin{prop}
\label{cor:CRLB MMV 2}Under the assumption that the error $\hat{\mathbf{H}}_{1}-\mathbf{H}_{1}$
consists of i.i.d., zero mean, Gaussian elements, the per-element
MSE of any unbiased estimator of $\mathbf{H}$ from $\hat{\mathbf{H}}_{1}$
that treats the rows of $\mathbf{H}$ as MMV, is lower bounded as
\begin{align}
\frac{1}{MN}\mathbb{E}(\|\hat{\mathbf{H}}-\mathbf{\mathbf{H}}\|^{2}) & \geq\frac{L^{2}\sigma^{2}(1+2N_{p})(1+2M)}{4MM_{p}N_{p}^{2}}\label{eq:seq. MMV CRLB detailed version}\\
 & \approx\frac{L^{2}\sigma^{2}}{M_{p}N_{p}}\text{ (for }N,M_{p}\gg1\text{).}\label{eq: seq. MMV CRLB}
\end{align}
\end{prop}
\begin{IEEEproof}
Note that treating the columns $\mathbf{H}_{1}$ as MMV results in
the estimation of $L(2+2N_{p})$ real-valued parameters. Following
the same lines as in the proof of Theorem \ref{thm: 2D CRLB}, it
can be shown that the per-element MSE of the estimate $\hat{\mathbf{H}}_{1}\in\mathbb{C}^{M\times N_{p}}$
obtained by processing $\mathbf{Y}$ is lower bounded as $\frac{1}{MN_{p}}\mathbb{E}(\|\hat{\mathbf{H}}_{1}-\mathbf{\mathbf{H}}_{1}\|^{2})\geq\frac{L(1+2N_{p})\sigma^{2}}{2M_{p}N_{p}}.$
Similarly, treating the rows of $\mathbf{H}$ as MMV results in the
estimation of $L(2+2M)$ real-valued parameters. The result of (\ref{eq:seq. MMV CRLB detailed version})
follows by the same lines as in the proof for the MSE bound of $\hat{\mathbf{H}}_{1}$,
under the assumption that $\hat{\mathbf{H}}_{1}=\mathbf{\mathbf{H}}_{1}+\mathbf{E}_{1}$,
where $\mathbf{E}_{1}\in\mathbb{C}^{M\times N_{p}}$ consists of i.i.d.,
zero mean Gaussian elements whose variance cannot be smaller than
$\frac{L(1+2N_{p})\sigma^{2}}{2M_{p}N_{p}}$.
\end{IEEEproof}
The above result suggests that the proposed channel estimator can
potentially achieve an MSE that is surprisingly only about $L/2$
times greater than the universal MSE bound of (\ref{eq:2D CRLB bound}),
when $L\geq2$. Interestingly, although Proposition \ref{cor:CRLB MMV 2}
is obtained under assumptions on the estimation error of $\hat{\mathbf{H}}_{1}$
that do not actually hold, the MSE formula of (\ref{eq:seq. MMV CRLB detailed version})
closely matches the actual MSE performance as will be demonstrated
in the numerical results of the next section.

\section{Numerical Results}

This section evaluates the MSE performance of the proposed channel
estimation scheme, described in Algorithm \ref{alg:proposed alg},
by means of Monte Carlo simulations. A system with $M=N=100$ was
considered, for which computation of the ANM-based channel estimation
of (\ref{eq:2D ANM problem formulation}) is practically infeasible.
The channel path gains were generated as i.i.d. complex Gaussian of
zero mean and variance $1/L$, resulting in an average received power
per subcarrier equal to $\sum_{l=0}^{L-1}\mathbb{E}(|c_{l}|^{2})=1$.
Path delays and AoAs were independently generated and uniformly distributed
in $[0,1/4]$ and $[0,1]$, respectively. This range of path delay
values corresponds to a maximum delay spread equal to $1/4$ the OFDM
symbol duration, which is a reasonable value in practical systems
\cite{Ahmadi LTE}. It is noted that \emph{no restrictions on the
separation of the paths in the delay-AoA domain were imposed}, unless
stated otherwise. The channel variance was set to $\sigma^{2}=0.1$,
for a per-subcarrier average SNR equal to $10$ dB. Targeting the
smallest possible pilot overhead $N_{p}$, full antenna observations
were considered, i.e., $M_{p}=M$. For any given $N_{p}$, MSE was
obtained by averaging over multiple independent realizations of $\mathbf{H}$,
$\mathbf{Z}$ and $\mathcal{N}_{p}$.

\begin{figure}[t]
\begin{centering}
\textsf{\includegraphics[width=1\columnwidth]{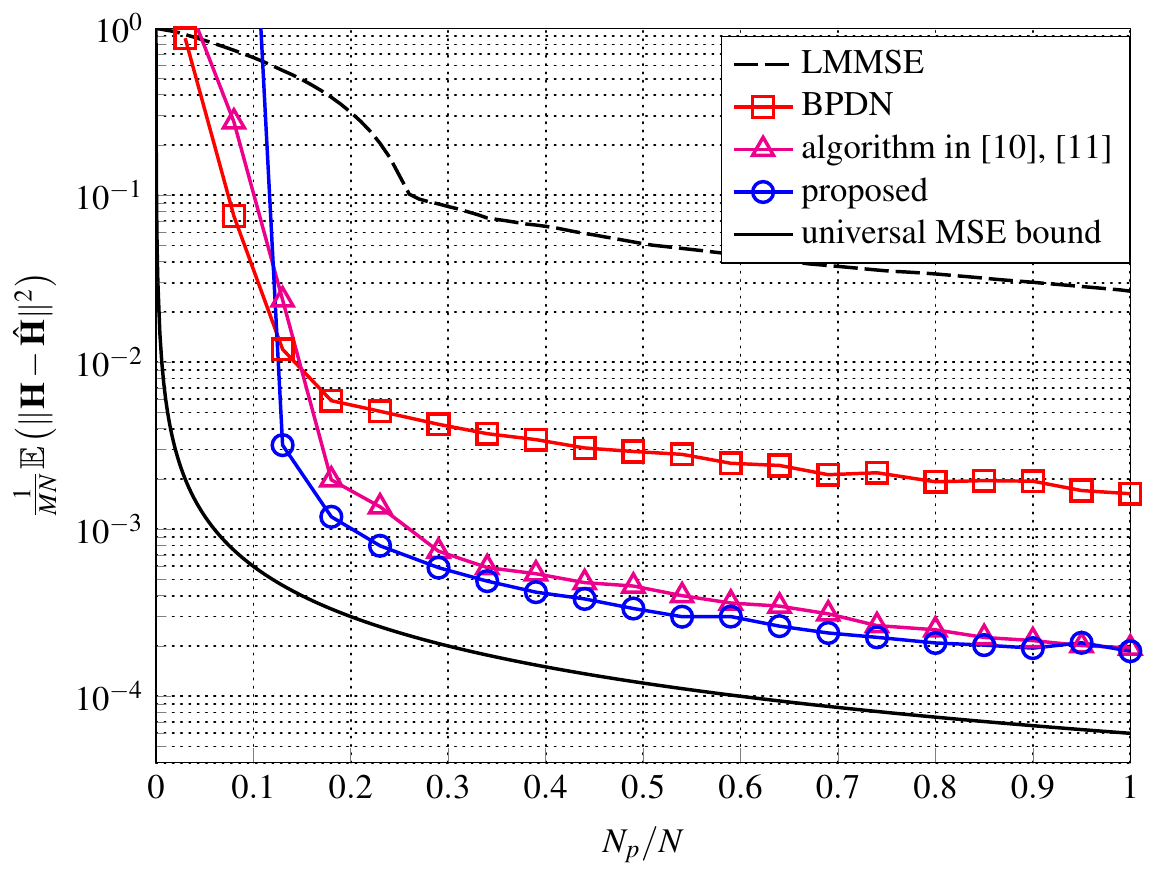}}
\par\end{centering}
\caption{\label{fig: MSE vs. N_p}MSE as a function of $N_{p}/N$ ($M=N=M_{p}=100,$
$\text{SNR}=10\text{ dB}$, $L=3$).}
\end{figure}

Figure \ref{fig: MSE vs. N_p} depicts the per-element MSE achieved
by the proposed algorithm as a function of the normalized pilot overhead
$N_{p}/N$. A sparse channel with $L=3$ paths was assumed. It can
be seen that excellent MSE performance is achieved that is more than
one order of magnitude smaller than the noise level with a pilot overhead
around $12\%$. In addition, the MSE performance closely follows the
universal bound of (\ref{eq:2D CRLB bound}) over all range of $N_{p}$
values, being only about $3$ times greater. Note that Proposition
\ref{cor:CRLB MMV 2} predicts an $L/2=1.5$ times greater MSE. 

For comparison purposes, the performance of the standard basis pursuit
denoising (BPDN) algorithm \cite{MathIntroToCS}, straightforwardly
modified to serve as channel estimator for this setting, is also shown.
This algorithm assumes that delays and AoAs take values over an equisampled
grid of their original domains and always assumes $\|\mathbf{Z}\|=\sqrt{N_{p}M_{p}\sigma^{2}}$,
which holds approximately by the law of large numbers. For this example,
$256$ samples where considered for both domains discretization, as
larger values resulted in high computational complexity with minimal
performance improvement. It can be seen that performance of BPDN is
significantly worse than the proposed algorithm performance except
in the regime of very small $N_{p}$ (but of high MSE as well). Also
shown is the performance of the conventional LMMSE estimator obtained
by a straightforward generalization of the approach in \cite{OFDM_chan_est_by_SVD}.
It can be seen that LMMSE performs poorly since the second order channel
statistics fail to capture the sparsity of the channel, resulting
in a minimum of $25\%$ pilot overhead to even achieve an MSE equal
to the noise level. 

The performance of the decoupled ANM estimator of \cite{ICASP 2017 second,ICASP 2017}
is also shown in Fig. \ref{fig: MSE vs. N_p}, operating by treating
$\mathbf{Y}$ as noiseless, extracting the (decoupled) AoA and delay
estimates as the $L$ largest harmonics, and obtaining a denoised
channel matrix estimate after coupling the AoA and delay estimates
using a procedure similar to the one in \cite{matrix pencil}. As
this algorithm demonstrated large sensitivity to path separability,
resulting in extremely poor estimates for channel realizations with
very closely spaced paths on the AoA-delay plane, its performance
is shown by averaging only over channel realizations satisfying (\ref{eq:2D separability condition})
with $d=1/N$. It can be seen that, even under this performance favorable
path separation assumption, the algorithm of \cite{ICASP 2017 second,ICASP 2017}
still performs worse than the proposed one (which effectively operates
with $d=0$), over almost all range of $N_{p}$. Note that the proposed
algorithm cannot resolve paths that are very closely spaced due to
fundamental estimation limits \cite{Stoica book} and inevitably results
in inaccurate path parameter estimates. However, even though inaccurate,
these estimates correspond to MMV atom sets over the AoA and angle
domains that can accurately represent the channel when considered
as MMV in these domains. This, in addition to the independent treatment
of the domains that requires no coupling of parameter estimates, results
in a robust algorithm performance even under channel realizations
with non-resolvable paths. 

\begin{figure}[t]
\begin{centering}
\textsf{\includegraphics[width=1\columnwidth]{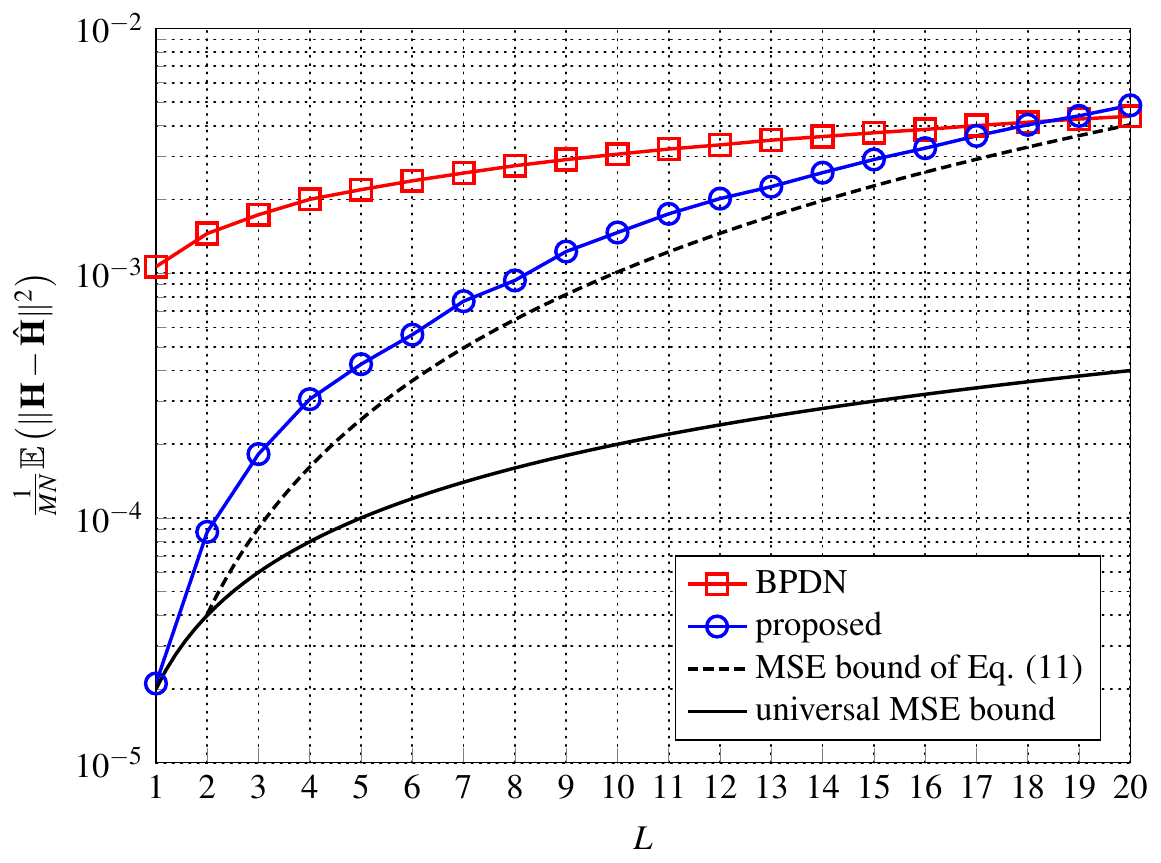}}
\par\end{centering}
\caption{\label{fig:MSE vs L}MSE as a function of $L$ ($M=N=N_{p}=M_{p}=100,$
$\text{SNR}=10\text{ dB}$).}
\end{figure}

Figure \ref{fig:MSE vs L} shows the MSE of the proposed algorithm
and BPDN as a function of $L$ with full observations, i.e., $N_{p}=N$,
$M_{p}=M$ (results are similar or $N_{p}<N$ and/or $M_{p}<M$).
It can be seen that the proposed algorithm outperforms BPDN for $L$
up to $18$, making it preferable for operations under sparse channels.
The MSE bound of (\ref{eq: seq. MMV CRLB}) is also shown (for $L\geq2)$,
where it can be seen that it serves as a very tight approximation
of the actual performance. Note that the performance of the proposed
algorithm scales as $L^{2}$, as predicted by (\ref{eq: seq. MMV CRLB}),
which eventually leads to the poorer performance compared to BPDN
for large $L$. However, for small $L$, performance is close to the
universal MSE bound of (\ref{eq:2D CRLB bound}).

\section{Conclusions}

A low-complexity sequential MMV ANM channel estimation algorithm was
proposed for wideband massive MIMO and its performance was analytically
characterized in terms of tight MSE bounds. It was demonstrated that
the algorithm can provide excellent performance in the regime of low-to-moderate
numbers of channel paths, without any restriction on their separability
in the AoA-delay plane. 

\section*{Acknowledgment}

This work has been performed in the framework of the Horizon 2020
project ONE5G (ICT-760809) receiving funds from the European Union.
The authors would like to acknowledge the contributions of their colleagues
in the project, although the views expressed in this contribution
are those of the authors and do not necessarily represent the project.
The work of G. Wunder was also supported by DFG grants WU 598/7-1
and WU 598/8-1 (DFG Priority Program on Compressed Sensing).

\appendices{}

\section{Proof of Theorem \ref{thm: 2D CRLB} \label{sec: 2D CRLB proof}}

For simplicity of exposition, the real-valued one-dimensional observation
model $\mathbf{y}=\mathbf{S}_{\mathcal{M}_{p}}\mathbf{h}+\mathbf{z}$
is considered first, where $\mathbf{h}=\mathbf{h}(\bm{\phi})\in\mathbb{R}^{M}$
is the vector to be estimated that depends continuously (but, otherwise,
arbitrarily), on a parameter vector $\bm{\phi}\in\mathbb{R}^{L}$,
$\mathbf{z}\in\mathbb{R}^{M_{p}}$ is a noise vector of i.i.d. Gaussian
elements of zero mean and variance $\sigma^{2}/2$, and and $\mathbf{S}_{\mathcal{M}_{p}}$
as defined in (\ref{eq: observations}) with $\mathcal{M}_{p}$ a
randomly and uniformly selected subset of $[M]$ with $M_{p}$ elements.
For a fixed $\mathcal{M}_{p}$, the covariance matrix of the error
$\hat{\mathbf{h}}-\mathbf{h}$ for any estimate $\hat{\mathbf{h}}$
is lower bounded as \cite{Kay Estimation theory}
\[
\mathbb{E}\left(\left(\hat{\mathbf{h}}-\mathbf{h}\right)\left(\hat{\mathbf{h}}-\mathbf{h}\right)^{H}\right)\succeq\nabla_{\bm{\phi}}\left(\mathbf{h}\right)\mathbf{I}^{-1}(\bm{\phi})\nabla_{\bm{\phi}}^{T}\left(\mathbf{h}\right),
\]
where the expectation is over noise statistics, $\nabla_{\bm{\phi}}\left(\mathbf{h}\right)\in\mathbb{R}^{M\times L}$
is the gradient of $\mathbf{h}$ w.r.t. to $\bm{\phi}$, and $\mathbf{I}(\bm{\theta})\in\mathbb{R}^{L\times L}$
is the Fisher information matrix for the parameter vector $\bm{\phi}$.
The latter is equal to \cite{Kay Estimation theory}
\begin{align*}
\mathbf{I}(\bm{\theta}) & =\frac{2}{\sigma^{2}}\nabla_{\bm{\phi}}^{T}(\mathbb{E}(\mathbf{y}))\nabla_{\bm{\phi}}(\mathbb{E}(\mathbf{y}))\\
 & =\frac{2}{\sigma^{2}}\nabla_{\bm{\phi}}^{T}(\mathbf{S}_{\mathcal{M}_{p}}\mathbf{h})\nabla_{\bm{\phi}}(\mathbf{S}_{\mathcal{M}_{p}}\mathbf{h})\\
 & =\frac{2}{\sigma^{2}}\nabla_{\bm{\phi}}^{T}(\mathbf{h})\mathbf{D}_{\mathcal{M}_{p}}\nabla_{\bm{\phi}}(\mathbf{h}),
\end{align*}
where $\mathbf{D}_{\mathcal{M}_{p}}\triangleq\mathbf{S}_{\mathcal{M}_{p}}^{T}\mathbf{S}_{\mathcal{M}_{p}}\in\mathbb{R}^{M\times M}$
is a diagonal matrix whose diagonal elements equal $[\mathbf{D}_{\mathcal{M}_{p}}]_{k,k}=1$,
if $k\in\mathcal{M}_{p}$, and $0$, if $k\in[M]\setminus\mathcal{M}_{p}$.
It follows that the per-element MSE is bounded as
\begin{align}
 & \frac{1}{M}\mathbb{E}(\|\hat{\mathbf{h}}-\mathbf{h}\|^{2})\\
= & \frac{1}{M}\text{tr}\left\{ \mathbb{E}\left(\left(\hat{\mathbf{h}}-\mathbf{h}\right)\left(\hat{\mathbf{h}}-\mathbf{h}\right)^{H}\right)\right\} \\
\geq & \frac{\sigma^{2}}{2M}\text{tr}\left\{ \nabla_{\bm{\phi}}(\mathbf{h})\left(\nabla_{\bm{\phi}}^{T}(\mathbf{h})\mathbf{D}_{\mathcal{M}_{p}}\nabla_{\bm{\phi}}(\mathbf{h})\right)^{-1}\nabla_{\bm{\phi}}^{T}(\mathbf{h})\right\} .\label{eq:MSE_conditioned_on_M}
\end{align}

A lower bound for the MSE can now be obtained by averaging the right-hand
side of (\ref{eq:MSE_conditioned_on_M}) with respect to the statistics
of $\mathcal{M}_{p}$. This is a non-tractable task in general, therefore,
looser lower bounds are pursued next. 

First note by a trivial application of the Cauchy-Schawrz inequality
that it holds $\sum_{k=0}^{L-1}a_{k}\geq L^{2}\left(\sum_{k=0}^{L-1}\frac{1}{a_{k}}\right)^{-1}$
for any sequence $\{a_{k}\}_{k=0}^{L-1}$ of strictly positive elements.
Also note that $\nabla_{\bm{\phi}}\left(\mathbf{h}\right)\left(\nabla_{\bm{\phi}}^{T}(\mathbf{h})\mathbf{D}_{\mathcal{M}_{p}}\nabla_{\bm{\phi}}(\mathbf{h})\right)^{-1}\nabla_{\bm{\phi}}^{T}\left(\mathbf{h}\right)$
is a positive semidefinite matrix of rank $L$ and let $\{\lambda_{k}\}_{k=0}^{L-1}$
denote its positive eigenvalues. Noting that $\{\lambda_{k}\}_{k=0}^{L-1}$
are also the eigenvalues of $\left(\nabla_{\bm{\phi}}^{T}(\mathbf{h})\mathbf{D}_{\mathcal{M}_{p}}\nabla_{\bm{\phi}}(\mathbf{h})\right)^{-1}\nabla_{\bm{\phi}}^{T}\left(\mathbf{h}\right)\nabla_{\bm{\phi}}\left(\mathbf{h}\right)$
\cite[Theorem 1.3.22]{Matrix Analysis}, it follows that
\begin{align}
 & \text{tr}\left\{ \nabla_{\bm{\phi}}\left(\mathbf{h}\right)\left(\nabla_{\bm{\phi}}^{T}(\mathbf{h})\mathbf{D}_{\mathcal{M}_{p}}\nabla_{\bm{\phi}}(\mathbf{h})\right)^{-1}\nabla_{\bm{\phi}}^{T}\left(\mathbf{h}\right)\right\} \nonumber \\
= & \text{tr}\left\{ \left(\nabla_{\bm{\phi}}^{T}(\mathbf{h})\mathbf{D}_{\mathcal{M}_{p}}\nabla_{\bm{\phi}}(\mathbf{h})\right)^{-1}\nabla_{\bm{\phi}}^{T}\left(\mathbf{h}\right)\nabla_{\bm{\phi}}\left(\mathbf{h}\right)\right\} \nonumber \\
= & \sum_{k=0}^{L-1}\lambda_{k}\nonumber \\
\geq & L^{2}\left(\sum_{k=0}^{L-1}\frac{1}{\lambda_{k}}\right)^{-1}\nonumber \\
= & L^{2}\left(\text{tr}\left\{ \left[\left(\nabla_{\bm{\phi}}^{T}(\mathbf{h})\mathbf{D}_{\mathcal{M}_{p}}\nabla_{\bm{\phi}}(\mathbf{h})\right)^{-1}\nabla_{\bm{\phi}}^{T}\left(\mathbf{h}\right)\nabla_{\bm{\phi}}\left(\mathbf{h}\right)\right]^{-1}\right\} \right)^{-1}\nonumber \\
= & L^{2}\left(\text{tr}\left\{ \mathbf{D}_{\mathcal{M}_{p}}\nabla_{\bm{\phi}}\left(\mathbf{h}\right)\left(\nabla_{\bm{\phi}}^{T}(\mathbf{h})\nabla(\mathbf{h})\right)^{-1}\nabla_{\bm{\phi}}^{T}\left(\mathbf{h}\right)\right\} \right)^{-1}.\label{eq:trace bound}
\end{align}

\noindent It follows from (\ref{eq:MSE_conditioned_on_M}) and (\ref{eq:trace bound})
that the per-element MSE, averaged over $\mathcal{M}_{p}$, is bounded
as
\begin{align*}
\mathsf{} & \frac{1}{M}\mathbb{E}(\|\hat{\mathbf{h}}-\mathbf{h}\|^{2})\\
\geq & \frac{L^{2}\sigma^{2}}{2M}\mathbb{E}\left(\text{tr}\left\{ \mathbf{D}_{\mathcal{M}_{p}}\nabla_{\bm{\phi}}\left(\mathbf{h}\right)\left(\nabla_{\bm{\phi}}^{T}(\mathbf{h})\nabla(\mathbf{h})\right)^{-1}\nabla_{\bm{\phi}}^{T}\left(\mathbf{h}\right)\right\} \right)^{-1}\\
\overset{(a)}{\geq} & \frac{L^{2}\sigma^{2}}{2M}\left(\text{tr}\left\{ \mathbb{E}\left(\mathbf{D}_{\mathcal{M}_{p}}\right)\nabla_{\bm{\phi}}\left(\mathbf{h}\right)\left(\nabla_{\bm{\phi}}^{T}(\mathbf{h})\nabla(\mathbf{h})\right)^{-1}\nabla_{\bm{\phi}}^{T}\left(\mathbf{h}\right)\right\} \right)^{-1}\\
\overset{(b)}{=} & \frac{L^{2}\sigma^{2}}{2M}\left(\frac{M_{p}}{M}\text{tr}\left\{ \nabla_{\bm{\phi}}\left(\mathbf{h}\right)\left(\nabla_{\bm{\phi}}^{T}(\mathbf{h})\nabla(\mathbf{h})\right)^{-1}\nabla_{\bm{\phi}}^{T}\left(\mathbf{h}\right)\right\} \right)^{-1}\\
= & \frac{L\sigma^{2}}{2M_{p}}
\end{align*}

\noindent where $(a)$ follows by application of Jenssen's inequality
and $(b)$ follows from $\mathbb{E}\left(\mathbf{D}_{\mathcal{M}_{p}}\right)=\frac{M_{p}}{M}\mathbf{I}$,
where $\mathbf{I}$ is the $M\times M$ identity matrix, which holds
by construction of $\mathcal{M}_{p}$. A straightforward but tedious
extension of this proof to the complex valued observation model of
(\ref{eq: observations}), noting that $\mathbf{H}$ depends on $4L$
real-valued parameters ($4$ parameters for each path: AoA, delay,
gain modulus and angle) leads to the result.

\end{document}